\documentclass[12pt]{article}
\usepackage{amsmath}
\usepackage{amssymb}
\usepackage{eufrak}

\begin{document}
\begin{center}
\section*{Fidelity and Wilson loop for quarks in confinement region}
\vskip 5mm V. I. Kuvshinov$^{1}$, P. V. Buividovich$^{2}$ \vskip
5mm {\small (1) {\it  JIPNR, National Academy of Science, Belarus,
220109 Minsk, Acad. Krasin str. 99, E-mail:  v.kuvshinov@sosny.bas-net.by}
\\
(2) {\it Belarusian State University, Minsk, F. Skoriny av. 4, E-mail: buividovich@tut.by} }
\end{center}
\vskip 5mm
\begin{abstract}
Connection between the stability of quantum motion in random
fields and quark confinement in QCD is investigated. The analogy
between the fidelity and the Wilson loop is conjectured, and the
fidelity decay rates for different types of quark motion are
expressed in terms of the parameters which are commonly used in
phenomenological and lattice QCD.
\end{abstract}

\small{PACS numbers: 12.38.Aw; 05.45.Mt} \\

\bigskip

The property of quark confinement in QCD is believed to be
determined by the presence of chaotic solutions in the spectrum of
Yang-Mills equations \cite{Sim1, Sim2}. In this case it is
important to investigate connection between the stability of quark
motion in random fields and the property of quark confinement.The
stability of quantum motion is usually described in terms of the
fidelity, and the confining properties are analyzed using the
Wilson loop. The aim of this paper is to demonstrate the
similarity between this two quantities in QCD and therefore to
reveal the analogy between the stability of quark motion and the
quark confinement.

One usually uses Wilson's area law for the Wilson loop as a litmus
test for quark confinement. Wilson loop is usually defined as the
trace of an averaged multiplicative integral over a closed contour
$C$:

\begin{equation}
\label{eq1} \hat {U}\left( {C} \right) = \overline {\hat
{P} \exp\left( {ig\int\limits_{C} {\left( {\hat {A}_{\mu} ^{0} +
\hat {A}_{\mu} ^{s}}  \right)dx^{\mu} }} \right)} ,\quad W\left(
{C} \right) = {\rm Tr} \left( {\hat {U}\left( {C} \right)} \right)
\end{equation}

\noindent where $\hat A_{\mu} ^{0}$ is a determined field, $\hat
A_{\mu} ^{s}$ is a random field and averaging is performed over
the ensemble of random fields $\hat {A}_{\mu} ^{s} $. The hat
symbol denotes operators in the colour space. If the ``area law''
holds for the Wilson loop $W\left( {C} \right)$, that is $W\left(
{C} \right) \sim \exp \left( - \sigma S_{c} \right)$ , $S_{c} $
being the minimal area of the surface spanned over the contour
$C$, quarks are said to be tied by a string with the constant
``tension'' $\sigma $ \cite{Sim1, Sim2}. In the case of stochastic
vacuum in QCD one usually chooses the curvature tensor of gauge
field $\hat {F}_{\mu \nu}  = \partial _{\mu}  \hat {A}_{\nu}  -
\partial _{\nu}  \hat {A}_{\mu}  - ig\left[ {\hat {A}_{\mu} ,\hat
{A}_{\nu} }  \right]_{ -}  $ as a random variable, because for
such a field gauge invariance is explicitly preserved. As the
stochastic vacuum should be colour neutral, here the random field
of curvature tensor $\hat {F}_{\mu \nu}  $ is assumed to be
statistically homogeneous with zero mean value and with
correlators proportional to identity in the colour space
\cite{Sim1}:

\begin{equation}
\label{eq2}
\begin{array}{l}
\overline {\hat {F}_{\mu \nu} }  = 0,       \quad g^{2}\overline
{\hat {F}_{\mu \nu } \left( {x_{1}} \right)\hat {F}_{\alpha \beta}
\left( x_{2}  \right)} = \hat {C}_{\mu \nu \alpha \beta}  \cdot
f\left( x_{1} -
x_{2} \right), \\
\quad f\left( 0 \right) = 1,\quad \hat {C}_{\mu \nu \alpha \beta}
= F^{2} \hat {I},\quad \int {d\xi ^{2}f\left( {\xi} \right)} =
l_{corr}^{2}
\end{array}
\end{equation}

For the stochastic vacuum the Wilson loop can be explicitly
calculated under conditions (\ref{eq2}) for the contour sizes
considerably exceeding the correlation length $l_{corr} $. For
topologically trivial fields the integral over the contour in
(\ref{eq1}) can be represented as the integral over the surface
spanned over the contour by applying the non-Abelian Stokes
theorem \cite{Sim1, Sim2, NAST}:

\begin{equation}
\label{eq3}
\hat {U}\left( {C} \right) = \overline {\hat {P}  \exp\left(
{ig\int\limits_{S\left( {C} \right)} {\tilde {F}_{\mu \nu} }  dS^{\mu \nu} }
\right)}
\end{equation}

\noindent
where $\tilde {F}_{\mu \nu}  = \hat {U}\left( {x,y} \right)\hat {F}_{\mu \nu
} \left( {y} \right)\hat {U}\left( {y,x} \right)$ is the shifted curvature
tensor, $\hat {U}\left( {x,y} \right) = \hat {P} \exp \left(
{ig\int\limits_{y}^{x} {\hat {A}_{\mu} }  dx^{\mu} } \right)$.

By applying the Van-Kampen expansion \cite{Kampen} to the integral
(\ref{eq3}) one can express it in terms of the accumulants of the
random field $\tilde {F}_{\mu \nu}$:

\begin{equation}
\label{KampenExpansion}
 W \left( C \right ) = \exp \left ( \sum_{k} i^{k} \Delta^{k}\left[ S \right] \right)
\end{equation}

\noindent where $\Delta^{k}\left[ S \right] = \frac{g^{k}}{k!}
\int_{S} d \sigma_{\mu_{1}\nu_{1}} \ldots \int_{S} d
\sigma_{\mu_{k}\nu_{k}} D_{\mu_{1}\nu_{1}\ldots\mu_{k}\nu_{k}}  $,
\\
$D_{\mu_{1}\nu_{1}\ldots\mu_{k}\nu_{k}} = {\rm Tr} \overline{\left (
\hat{F}_{\mu_{1}\nu_{1}} \ldots  \tilde{F}_{\mu_{k}\nu_{k}}
\right)} $. An efficient estimation for accumulants in
(\ref{KampenExpansion}) can be obtained for sufficiently large
contours after taking (\ref{eq2}) into account:

\begin{equation}
\label{Estimation} \Delta^{k}\left[ S \right]\approx
\frac{g^{k}}{k!}F^{k}l_{corr}^{2 (k - 1)}S
\end{equation}

The contribution of the second-order accumulant dominates in the
Wilson loop if the condition $\Delta^{k}\left[ S \right] \ll
\Delta^{2}\left[ S \right]$, $ k>2$ holds, or, according to the
estimation (\ref{Estimation}):

\begin{equation}
\label{GaussDominance}
g\sqrt {\overline {\hat {F}^{2}}} \cdot
l_{corr}^{2} \ll 1
\end{equation}
This condition is called the condition of the
gaussian-dominated vacuum in QCD \cite{Sim1, Sim2}. Under the
assumption of the gaussian-dominated vacuum the final expression
for the Wilson loop (\ref{eq1}) is:

\begin{equation}
\label{eq9} W\left( {C} \right)\sim \exp \left( - \Delta^{2}\left[
S \right]\right) \sim \exp \left( - \frac{g^{2}}{2} l_{corr}^{2}
F^{2} S \right )
\end{equation}

Thus the Wilson area law for the Wilson loop holds, and the
gaussian-dominated stochastic vacuum possesses confining
properties. The string tension $\sigma$ is also obtained from
(\ref{eq9}): $\sigma = \frac{g^{2}}{2} l_{corr}^{2} F^{2}$.
\bigskip

Similar to confinement and Wilson loop in QCD, the stability of
quantum motion is commonly described in terms of fidelity decay.
Fidelity is usually defined as the scalar product of the state
vectors of perturbed and unperturbed systems \cite{Kuvshinov,
Prosen}. For the purposes of semiclassical analysis of quark
motion without taking spin into account it is convenient to define
the fidelity as the scalar product of the state vectors in the
colour space:

\begin{equation}
\label{eq10}
f = \overline { \langle f_{1}  | f_{2}  \rangle } ,
\quad 
| f_{1} \rangle,
| f_{2} \rangle \in \mathbb{C}^{3},
\quad
| f_{1} | = | f_{2} | = 1
\end{equation}

\noindent where averaging is performed over random perturbations
of the system. According to the standard treatment of state
vectors in quantum theory it is more natural to average the square
of the absolute value of the fidelity $|f|^{2}$. However for the
estimation of the fidelity decay rate such averaging is possible.
One can naturally expect the absolute value of the fidelity to
decay, the decay rate being approximately equal for almost all
``typical'' implementations of the random field. Exponential decay
of the absolute value of the fidelity for the fidelity values
close to unity is described by the linear term in the Taylor
expansion: $f(t) \approx 1 - \alpha t$, $\alpha$ being the
fidelity decay rate and $t$ being some parameter on which the
fidelity depends. Other possible mechanism for the decay of the
averaged fidelity is due to the randomly changing phase of the
fidelity. However for the estimation of the fidelity decay rate
this mechanism can be neglected, as it is possible to show that
the fidelity decay in this case is described by the factor $1 -
(\alpha t)^{2}$, and $\alpha t$ is small for the fidelity values
close to unity. In this case the averaged fidelity should be close
enough to the absolute value of the fidelity for the ``typical''
implementations of random variables \cite{RMT,Stat}.

The first interesting case is the motion of coloured quark in the
different paths $\gamma _{1} $ and $\gamma _{2} $ in chaotic
environment. The paths start from the point $x$ and join in the
point $y$. In the point $x$ the state vector is $\left| {f_{0}}
\right\rangle $. In the limit of very massive quarks \cite{Sim1,
Sim2} the evolution of the state vector in the colour space is
described by the multiplicative integral introduced in
(\ref{eq1}):

\begin{equation}
\label{eq11}
\begin{array}{l}
\left| {f_{1}}  \right\rangle = \hat {P} \exp\left(
{ig\int\limits_{\gamma _{1}}  {\hat {A}_{\mu}  dx^{\mu} }}
\right)\left| {f_{0}}  \right\rangle = \hat {U}\left( {\gamma
_{1}}  \right)\left| {f_{0}}  \right\rangle , \\
\quad \left| {f_{2}} \right\rangle = \hat {P} \exp\left(
{ig\int\limits_{\gamma _{2}} {\hat {A}_{\mu}  dx^{\mu} }}
\right)\left| {f_{0}} \right\rangle = \hat {U}\left( {\gamma _{2}}
\right)\left| {f_{1}} \right\rangle
\end{array}
\end{equation}
The operators $\hat {U}\left( {\gamma _{1}}  \right)$
and $\hat {U}\left( {\gamma _{2}}  \right)$ are unitary because
$\hat {A}_{\mu}  $ is hermitan. Taking this into account, one can
rewrite the expression for the fidelity:

\begin{equation}
\label{eq12}
f = \left\langle {f_{0}}  \right|\overline {\hat
{U}\left( {\gamma _{1}} \right) \cdot \hat {U}^{ +} \left( {\gamma
_{2}}  \right)} \left| {f_{0}} \right\rangle = \left\langle
{f_{0}}  \right|\overline {\hat {U}\left( {\gamma _{1} \bar
{\gamma} _{2}}  \right)} \left| {f_{0}}  \right\rangle
\end{equation}

\noindent where $\gamma _{1} \bar {\gamma} _{2} $ is the path
obtained by travelling from the point $x$ to the point $y$ in the
path $\gamma _{1} $ and back to the point $x$ in the path $\gamma
_{2} $, $\hat {U}\left( {\gamma _{1} \bar {\gamma} _{2}}  \right)
= \hat {P} \exp\left( {ig\int\limits_{\gamma _{1} \bar {\gamma}
_{2}}  {\hat {A}_{\mu}  dx^{\mu} }}  \right)$. The averaging in
(\ref{eq12}) is performed over the random field of the curvature
tensor of the gauge field. It is evident that the fidelity is
directly related to the Wilson operator $\hat {U}\left( {C}
\right)$ introduced in (\ref{eq1}). The Van-Kampen decomposition
can be again applied for the estimation of the fidelity
(\ref{eq12}) under the assumption of gaussian-dominated
colour-neutral vacuum. The final result is:

\begin{equation}
\label{eq13} f \sim \exp \left( - \frac{g^{2}}{2} l_{corr}^{2}
F^{2} S \right )
\end{equation}
The error of this estimation is approximately equal to
those of the estimation of the Wilson loop, which does not exceed
few percent \cite{Sim2}. Thus for the gauss-dominated stochastic
vacuum the fidelity for the quark moving in different paths decays
exponentially with the area of the surface spanned over the paths,
the decay rate being equal to the string ``tension'' $\sigma $.
This hints at the close connection between the stability of quark
motion and quark confinement.

Another possible situation, which is more close to the standard
treatment of the fidelity, is realized when $\gamma _{1} $ and
$\gamma _{2} $ are two random paths which are very close to each
other. The corresponding expression for the fidelity is similar to
(\ref{eq12}), but now the averaging is performed with respect to
all random paths which are close enough. The final result for the
averaged path-ordered integral is obtained in the way similar to
(\ref{eq3}) - (\ref{eq9}), but in this case the field variables
are regarded as predetermined. The final expression for the
fidelity in this case is:

\begin{equation}
\label{eq14} f = \left\langle {f_{0}}  \right|\hat {P} \exp\left( {
- \frac{{g^{2}l_{corr} }}{{2}} \int\limits_{\gamma _{1}} {\tilde
{F}_{\chi \alpha}  \tilde {F}_{\nu \beta}  n^{\chi}  \overline
{\delta x^{\alpha} \delta x^{\beta} } dx^{\nu} }} \right)\left|
{f_{0}} \right\rangle
\end{equation}

\noindent where $\delta x^{\alpha} $ is the deviation of the path
$\gamma _{2} $ from the path $\gamma _{1} $, $n^{\chi}$ is the
four-dimensional velocity. A rough estimation of the fidelity
(\ref{eq14}) is:

\begin{equation}
\label{eq15}
f \sim \exp\left( { - g^{2}l_{corr} \cdot F^{2} \cdot
\overline {\delta x^{2}} \cdot L} \right)
\end{equation}

where $F^{2}$ in this estimation is the trace of the square of the
curvature tensor $\hat {F}_{\mu \nu} $, $l_{corr}$ is the
correlation length of quark path expressed in terms of world line
length, and the length $L$ characterizes the length of the average
path both in time and space. For example, if the average path is
parallel to the time axis in the Minkowski space, the quark moves
randomly around some point in three-dimensional space. The
fidelity in this case decays exponentially with time, as would be
expected. \bigskip

Thus the fidelity decay is related to the mechanism of quark
confinement, at least for the simple model of gaussian-dominated
stochastic vacuum. The stronger the quarks are coupled to each
other, the greater is the fidelity decay rate. The exact proofs
and the consistent treatment of this phenomena in quantum field
theory require further investigations.


\begin{thebibliography}{9}
\bibitem{Sim1} Yu. A. Simonov, {\em Uspekhi Fizicheskih Nauk \/} 4 (1996).
\bibitem{Sim2} D.S. Kuz'menko, Yu. A. Simonov, V. I. Shevchenko, {\em Uspekhi Fizicheskih Nauk\/} 1 (2004).
\bibitem{NAST} Ying Chen, Bing He, Ji-Min Wu, {\em arXiv:hep-th/0006104 v2\/}.
\bibitem{Kampen} N. G. van Kampen, {\em Physica\/} 74, 215 (1974).
\bibitem{Kuvshinov} V.I. Kuvshinov, A.V. Kuzmin, {\em Physics Letters A\/} 316 (2003).
\bibitem{Prosen} T. Prosen, M. Znidaric, {\em J. Phys. A: Math. Gen.\/} 34 (2001) L681-L687.
\bibitem{RMT} M. Mehta, {\em Random Matrices\/}, 2nd edition, Academic
Press Inc., Boston, 1991.
\bibitem{Stat} R. Durett, {\em Probability: Theory and examples\/}, 2nd
edition, Duxbury Press, 1996.
\end{thebibliography}
\end{document}